\documentclass[12pt]{article}
\input{epsf.sty}
\def\fo{\hbox{{1}\kern-.25em\hbox{l}}}

\def\slashchar#1{\setbox0=\hbox{$#1$}           
   \dimen0=\wd0                                 
   \setbox1=\hbox{/} \dimen1=\wd1               
   \ifdim\dimen0>\dimen1                        
      \rlap{\hbox to \dimen0{\hfil/\hfil}}      
      #1                                        
   \else                                        
      \rlap{\hbox to \dimen1{\hfil$#1$\hfil}}   
      /                                         
   \fi}                                         %

\def\beq{\begin{equation}}
\def\eeq{\end{equation}}
\def\eq{\end{equation}}
\def\to{\rightarrow}

\def\mEt{\mbox{${\hbox{$E$\kern-0.6em\lower-.1ex\hbox{/}}}_T$}\, } 
\def\mE{\mbox{${\hbox{$E$\kern-0.6em\lower-.1ex\hbox{/}}}$}\, } 

\def\bsg{\ifmmode B\to X_s\gamma\else $B\to X_s\gamma$\fi}
\def\bsll{\ifmmode B\to X_s\ell^+\ell^-\else $B\to X_s\ell^+\ell^-$\fi}
\def\bstt{\ifmmode B\to X_s\tau^+\tau^-\else $B\to X_s\tau^+\tau^-$\fi}
\def\shat{\ifmmode \hat{s}\else $\hat{s}$\fi}

\newcommand{\newc}{\newcommand}
\def\womp#1{#1}
\def\wompit#1{#1}

\newc{\lcal}{\int {\cal L}dt}
 
\newc{\LSP}{{\chi^0_1}}
\newc{\stauR}{{\tilde \tau_R}}
\newc{\stau}{{\tilde \tau_1}}
\newc{\mstop}{m_{\tilde{t}}}
\newc{\mHpm}{m_{H^\pm}}
\newc{\gsim}{\lower.7ex\hbox{$\;\stackrel{\textstyle>}{\sim}\;$}}
\newc{\lsim}{\lower.7ex\hbox{$\;\stackrel{\textstyle<}{\sim}\;$}}
\newc{\ie}{{\it i.e.}}          
\newc{\etal}{{\it et al.}}
\newc{\eg}{{\it e.g.}}          
\newc{\kev}{\hbox{\rm\,keV}}            
\newc{\mev}{\hbox{\rm\,MeV}}            
\newc{\gev}{\hbox{\rm\,GeV}}            
\newc{\tev}{\hbox{\rm\,TeV}}
\newc{\xpb}{\hbox{\rm\, pb}}
\newc{\xfb}{\hbox{\rm\, fb}}

\newc{\mtop}{m_t}
\newc{\mbot}{m_b}
\newc{\mz}{m_Z}
\newc{\mw}{M_W}
\newc{\alphasmz}{\alpha_s(m_Z^2)}
\newc{\swsq}{\sin^2\theta_W}
\newc{\tw}{\tan\theta_W}
\newc{\cw}{\cos\theta_W}
\newc{\sw}{\sin\theta_W}
\newc{\BR}{\hbox{\rm BR}}
\newc{\zbb}{Z\to b\bar}
\newc{\Gb}{\Gamma (Z\to b\bar b)}
\newc{\Gh}{\Gamma (Z\to \hbox{\rm hadrons})}
\newc{\rbsm}{R_b^\hbox{\rm sm}}
\newc{\rbsusy}{R_b^\hbox{\rm susy}}
\newc{\drb}{\delta R_b}

\newc{\sgn}{\mbox{sgn}}
\newc{\tbeta}{\tan\beta}
\newc{\uL}{{\tilde u_L}}
\newc{\uR}{{\tilde u_R}}
\newc{\cL}{{\tilde c_L}}
\newc{\cR}{{\tilde c_R}}
\newc{\tL}{{\tilde t_L}}
\newc{\tR}{{\tilde t_R}}
\newc{\dL}{{\tilde d_L}}
\newc{\dR}{{\tilde d_R}}
\newc{\sL}{{\tilde s_L}}
\newc{\sR}{{\tilde s_R}}
\newc{\bL}{{\tilde b_L}}
\newc{\bR}{{\tilde b_R}}
\newc{\eL}{{\tilde e_L}}
\newc{\eR}{{\tilde e_R}}
\newc{\mhp}{m_{H^\pm}}
\newc{\mhalf}{m_{1/2}}
\newc{\emt}{{e/\mu /\tau}}

\newc{\lR}{\tilde{l}_R}
\newc{\lL}{\tilde{l}_L}
\newc{\nL}{\tilde{\nu}_L}
\newc{\na}{\chi^0_1}
\newc{\nb}{\chi^0_2}
\newc{\nc}{\chi^0_3}
\newc{\nd}{\chi^0_4}
\newc{\ca}{\chi^{\pm}_1}
\newc{\cb}{\chi^{\pm}_2}
\newc{\camp}{\chi^\mp_1}
\newc{\cbmp}{\chi^\mp_1}
\newc{\capos}{\chi^{+}_1}
\newc{\caneg}{\chi^{-}_1}
\newc{\phit}{\phi_t}
\newc{\phib}{\phi_b}
\newc{\phiew}{\phi_{ew}}
\newc{\htz}{h^0_t}
\newc{\hbz}{h^0_b}
\newc{\hewz}{h^0_{ew}}
\newc{\hsmz}{h^0_{sm}}
\newc{\huz}{h^0_u}
\newc{\hsusyz}{h^0_{susy}}

\def\NPB#1#2#3{Nucl. Phys. B {\bf #1}, #3 (19#2)}
\def\PLB#1#2#3{Phys. Lett. B {\bf #1}, #3 (19#2)}

\def\PRD#1#2#3{Phys. Rev. D {\bf #1}, #3 (19#2)}
\def\PRL#1#2#3{Phys. Rev. Lett. {\bf#1}, #3 (19#2)}
\def\PRT#1#2#3{Phys. Rep. {\bf#1}, #3 (19#2)}

\def\beq{\begin{equation}}
\def\eeq{\end{equation}}
\def\bea{\begin{eqnarray}}
\def\eea{\end{eqnarray}}
\catcode`@=11
\long\def\@caption#1[#2]#3{\par\addcontentsline{\csname
  ext@#1\endcsname}{#1}{\protect\numberline{\csname
  the#1\endcsname}{\ignorespaces #2}}\begingroup
    \small
    \@parboxrestore
    \@makecaption{\csname fnum@#1\endcsname}{\ignorespaces #3}\par
  \endgroup}
\catcode`@=12

\def\jfig#1#2#3{
 \begin{figure}
 \centering
 \epsfysize=3.0in
 \hspace*{0in}
 \epsffile{#2}
 \caption{#3}
 \label{#1}
 \end{figure}}
\def\bfig#1#2#3{
 \begin{figure}
 \centering
 \epsfysize=4.5in
 \hspace*{0in}
 \epsffile{#2}
 \caption{#3}
 \label{#1}
 \end{figure}}
\def\sfig#1#2#3{
 \begin{figure}
 \centering
 \epsfysize=2.5in
 \hspace*{0in}
 \epsffile{#2}
 \caption{#3}
 \label{#1}
 \end{figure}}

\begin{document}
\begin{titlepage}

\begin{flushleft}
\end{flushleft}
\begin{flushright}
hep-ph/9904378 \\
CERN-TH/99-104
\end{flushright}

\vspace{1cm}

\vspace*{15.3cm}

\begin{flushleft}
CERN-TH/99-104 \\
April 1999\\
\end{flushleft}

\vspace*{-17.0cm}



\huge
\begin{center}
{\Large\bf
Phenomenological Consequences of \\ 
Supersymmetry with Anomaly-Induced Masses}
\end{center}

\large

\vspace{.15in}
\begin{center}

Tony Gherghetta, Gian F.~Giudice, and James D.~Wells

\small

\vspace{.1in}
{\it CERN, Theory Division, CH-1211 Geneva 23, Switzerland \\}

\end{center}
 
 
\vspace{0.15in}
 
\begin{abstract}

In the supersymmetric standard model there exist pure gravity 
contributions to the soft mass parameters which arise 
via the superconformal anomaly.
We consider the low-energy phenomenology with a
mass spectrum dominated by the anomaly-induced contributions.
In a well-defined
minimal model we calculate electroweak symmetry breaking parameters,
scalar masses, and the full one-loop splitting of the degenerate
Wino states.
The most distinctive features are gaugino masses
proportional to the corresponding gauge coupling beta-functions,
the possibility of a Wino as the lightest supersymmetric particle, 
mass degeneracy of sleptons, and a
very massive gravitino.  
Unique signatures at high-energy colliders include 
dilepton and single lepton
final states, accompanied by missing energy
and displaced vertices. We also point out that
this scenario has the cosmological advantage of ameliorating the gravitino
problem.
Finally, the primordial gravitino decay can produce a relic density of
Wino particles close to the critical value.

\end{abstract}

\bigskip

\end{titlepage}

\baselineskip=18pt
\setcounter{footnote}{1}
\setcounter{page}{2}
\setcounter{figure}{0}
\setcounter{table}{0}
\vfill
\eject

\section{Introduction}
\medskip

Supersymmetry provides a promising 
solution to the gauge hierarchy problem afflicting the standard model 
(SM). However, it is clear that supersymmetry must be 
broken at low energies. The specific mechanism for transmitting
supersymmetry breaking effects is important in determining the
low-energy experimental signatures. Currently, there are two known ways
that supersymmetry breaking effects appear in the low-energy Lagrangian.
In gravity-mediated scenarios~\cite{grav}, 
supersymmetry is broken in a hidden sector 
and transmitted gravitationally to the observable sector fields. While
this scenario is elegant and simple, it suffers from the supersymmetric
flavor problem. Alternatively, in gauge-mediated 
scenarios~\cite{gaugemed}, supersymmetry 
breaking is transmitted via gauge forces and this scenario provides
an appealing solution to the supersymmetric flavor problem. Both of these
alternative scenarios have distinct experimental signatures.

We consider a third scenario for 
transmitting supersymmetry breaking to the observable sector.
In this scenario, rescaling anomalies in the supergravity Lagrangian
give rise to soft mass parameters for the observable sector 
fields~\cite{randall,giudice}. Unlike
the gravity-mediated or gauge-mediated scenarios, these anomaly 
contributions will always be present if supersymmetry is broken.
We will refer to the case in which the anomaly-induced masses 
are dominant as the anomaly-mediated supersymmetry breaking (AMSB) scenario.
In this scenario the gaugino mass is proportional to the corresponding 
gauge beta function while the scalar masses (and $A$-terms) depend on the 
anomalous dimensions of the corresponding scalar fields. One of the 
distinctive features of the AMSB scenario is the gaugino mass spectrum, with
the Wino being the lightest supersymmetric particle. Similarly, the
squark mass spectrum is unique but unfortunately the slepton mass spectrum 
is tachyonic. This can be cured by adding a positive, non-anomaly mediated 
contribution~\cite{randall}.  Some phenomenological consequences of this
scenario have been recently presented in ref.~\cite{feng}. A different and
very interesting approach to cure the tachyonic mass spectrum problem has been 
suggested in ref.~\cite{pr}.

A distinctive feature of AMSB is that the gravitino is much heavier
than the gauginos and squarks. This is cosmologically attractive
because the gravitino problem can be ameliorated. 
Moreover, gravitino decays can produce a
present Wino energy density close to the critical value. The neutral
Wino is therefore a good dark-matter candidate, in spite of its negligible
thermal relic density.

\section{The anomaly-induced mass spectrum}

The anomaly-induced soft terms~\cite{randall,giudice} are always present 
in a broken supergravity theory, regardless of the specific form of
the couplings between the hidden and observable sectors. They are linked
to the existence of the superconformal anomaly. Indeed they explicitly
arise when one tries to eliminate from the relevant Lagrangian the
supersymmetry-breaking auxiliary background field by making a suitable
Weyl rescaling of the superfields in the observable sector. Their origin
has been discussed from various point of views in 
refs.~\cite{randall,giudice,pr}. Here we give a heuristic
derivation of the
essential results, and make some comments on their phenomenological 
relevance.

The effect of supersymmetry breaking can be described by a flat-space
chiral superfield $\Phi$, with background value
\beq
   \Phi =1-m_{3/2} \theta^2.
\label{phiback}
\eeq
This field acts as a compensator of the super-Weyl transformation. In other
words, by choosing suitable couplings of $\Phi$ to the observable fields,
the theory is made superconformal invariant.

Let us consider a supersymmetric gauge theory with no mass parameters
at the classical level. This does not appear at first sight to be relevant
to the minimal
supersymmetric model which contains a mass term -- the Higgs mixing mass
$\mu$ -- seemingly even in the limit of exact supersymmetry. Actually,
the $\mu$ term can be viewed as an effect of 
supersymmetry breaking~\cite{gmas},
and therefore we set it to zero for the moment. Mechanisms for generating
$\mu$ in AMSB scenarios have been discussed in refs.~~\cite{randall,giudice,pr}.
At the quantum level, there is always the need to introduce a mass parameter,
which is the renormalization scale $\mu$ (not to be confused with the
Higgs mixing parameter). In the presence of a compensator field $\Phi$
for super-Weyl transformations, it is natural to expect that the 
renormalization scale $\mu$ is promoted to a superfield, according to
\beq
\mu \to \mu / \sqrt{\Phi^\dagger \Phi} .
\label{cont}
\eeq
The replacement of $\Phi$ with its background value given in Eq.~(\ref{phiback})
generates a specific set of supersymmetry-breaking terms.

The simplest way to obtain the form of the supersymmetry-breaking terms is
to employ the technique developed in ref.~\cite{gr}. The main idea is that
when certain parameters of a supersymmetric theory are ``analytically
continued" into superspace, the renormalization-group (RG) flow of the
modified theory is completely determined by the properties of the original
theory. In particular, if a parameter is continued into a 
supersymmetry-breaking background field, the RG properties of the exact
supersymmetric theory determine the form of the soft terms. The
prescription given in Eq.~(\ref{cont}) is a specific example of such a
continuation. We can then make use of the general expressions of the
gaugino masses $M_\lambda$, scalar masses $m_{\tilde Q}$, and 
trilinear couplings $A_{Q_i}$ in terms of derivatives of
the field wave-functions~\cite{gr},
\bea
M_\lambda &=&-\frac{1}{2} \left. \frac{\partial \ln S}{\partial \ln \Phi}
\right|_0 F_\Phi \\
m_{\tilde Q}^2&=& -\left. \frac{\partial^2 \ln Z_Q}{\partial \ln \Phi
\partial \ln \Phi^\dagger}
\right|_0 F_\Phi^\dagger F_\Phi \\
A_{Q_i} &=& \left. \frac{\partial \ln Z_{Q_i}}{\partial \ln \Phi}
\right|_0 F_\Phi .
\eea
The symbol ``$|_0$" denotes setting to zero the Grassmann coordinates,
$\theta =\bar \theta =0$. Here $S$ and $Z_Q$ are the gauge and matter field
wave-functions, with $S$ related to the gauge coupling constant
by Re$(S)|_0=g^{-2}/4$. Using Eq.~(\ref{cont}) and $F_\Phi =-m_{3/2}$, see
Eq.~(\ref{phiback}), we obtain
\bea
    M_\lambda &=& 
-\frac{g^2}{2}\frac{d g^{-2}}{d \ln \mu} m_{3/2}
=
{\beta_g\over g} m_{3/2} 
\label{gaugdef}
\\
  m_{\tilde Q}^2&=&-{1\over 4}
\frac{d^2 \ln Z_Q}{d(\ln \mu)^2} m_{3/2}^2=
-{1\over 4}
      \left({\partial\gamma\over\partial g}\beta_g +
       {\partial\gamma\over\partial y}\beta_y\right)m_{3/2}^2 
\label{squarkdef}\\
   A_{y}&=&\frac{1}{2}
    \sum_i \frac{d\ln Z_{Q_i}}{d\ln \mu} m_{3/2}=
        -{\beta_y\over y} m_{3/2} .\label{adef}
\eea
Here the sum $\sum_i$ extends over the fields involved in the Yukawa
superpotential term with coupling constant $y$, and
we have used the renormalization group functions
$\gamma(g,y)\equiv {d \ln Z/d\ln\mu}$, $\beta_g(g,y)\equiv 
{d g/d\ln\mu}$, and $\beta_y(g,y)\equiv {d y/d\ln\mu}$.

\subsection{Features of the anomaly-induced soft terms}

The soft terms in Eqs.~(\ref{gaugdef})--(\ref{adef}) are determined by
the anomalous dimensions of the fields or, in other words, by the 
violation of the Weyl symmetry in the quantum theory given by the conformal
anomaly. Indeed, the supergravity prescription in Eq.~(\ref{cont}) is
sufficient to determine the complete form of the soft terms, by means
of the technique of ref.~\cite{gr}.

The form of the soft terms in Eqs.~(\ref{gaugdef})--(\ref{adef})
is particularly interesting because it is invariant under RG 
transformations. This means that the analytic continuation into
superspace given by Eq.~(\ref{cont}) defines a consistent RG trajectory
for the soft terms. The phenomenological appeal of this form of the soft
terms resides precisely in this crucial property. In particular, it
entails a large degree of predictivity, since all soft terms can be
computed from known low-energy SM parameters and a single mass scale,
$m_{3/2}$. Also, it leads to robust predictions, since the RG invariance
guarantees complete insensitivity of the soft terms from ultraviolet
physics. As demonstrated with specific examples in ref.~\cite{giudice},
heavy states do not affect the low-energy parameters in 
Eqs.~(\ref{gaugdef})--(\ref{adef}), since their effects in the 
beta-functions and threshold corrections exactly compensate each other.
This means that the gaugino mass prediction in 
Eq.~(\ref{gaugdef}) is valid irrespective of the GUT gauge group in which
the SM may or may not be embedded. However, exceptions to ultraviolet
insensitivity appear in the presence of gauge singlet superfields~\cite{pr}.

The insensitivity from ultraviolet physics not only leads to robust
predictivity, but also provides a solution to the supersymmetric flavor
problem. Indeed the unknown physics which breaks the flavor symmetry at a 
high-energy scale $\Lambda_F$ and determines the Yukawa couplings does not
leave any visible trace in the anomaly-mediated
soft terms. Recall that in gauge mediation
the flavor problem is solved by making the soft terms insensitive to any
physics above the messenger scale $M$. The parameter $M$ is unknown, and
is chosen such that $M<\Lambda_F$. The soft terms vanish above the
scale $M$ and therefore their low-energy values are finite and
have a logarithmic dependence on $M$. In contrast, in anomaly mediation
the soft terms do not vanish at any scale (below the Planck mass $M_{P}$), 
but their
values at low energies are not influenced by physics at any intermediate
scale.

In order to preserve the attractive properties of the anomaly-mediated
soft terms, we have to make sure that other forms of communication of
supersymmetry breaking to the observable sector do not give larger
contributions. In ordinary gravity mediation, one makes use of 
tree-level supersymmetry-breaking communication which, in general, dominates
over the loop effects of anomaly mediation. If there are no gauge-singlet
superfields with scalar vacuum expectation value of order $M_{P}$,
then the theory does not contain operators of the form
\beq
   \int d^2\theta {X\over M_P} 
{\rm Tr}{\cal W}^\alpha{\cal W}_\alpha +
  {\rm h.c.},
\eeq
where $X$ is the Goldstino superfield. Gaugino masses are only generated
by higher-dimensional operators and are at best 
of order $m_{3/2}^{3/2}/M_P^{1/2}$. 
In particular, this is in general
true in theories with dynamical supersymmetry breaking. In this case, the
anomaly-mediated effects give the dominant contributions to gaugino
masses~\cite{giudice}.

It appears at first difficult to forbid or suppress
tree-level gravity contributions to scalar masses, which are obtained by
couplings in the K\"ahler potential between
visible sector fields $Q$ and the Goldstino multiplet $X$,
\beq
\label{directcoupling}
   \int d^4\theta {1\over M_P^2} X^\dagger X Q^\dagger Q.
\eeq
However, the suppression is possible if the
K\"ahler potential has the specific structure
\beq
   K=-3 M_P^2\ln(1-{f_{\rm vis}\over 3 M_P^2} -{f_{\rm hid}\over 3 M_P^2}),
\label{kaler}
\eeq
where $f_{\rm vis}$ and $f_{\rm hid}$ are functions of only
visible and hidden fields, respectively.
This structure could be the result of the 
underlying fundamental theory such as string theory. However, 
it is not clear how such a special form of the K\"ahler 
potential can be stable against radiative corrections.

A very interesting possibility,
pointed out in ref.~\cite{randall}, 
is that the supersymmetry-breaking and visible sectors reside on different
branes embedded into a higher-dimensional space and
separated by a sufficiently large distance. In this case, the 
structure in Eq.~(\ref{kaler}) is guaranteed by the geometry and not by
a symmetry. 
Thus, all the low-energy soft
parameters will arise from anomaly-induced effects.

Unfortunately, it turns out that the pure scalar mass-squared anomaly 
contribution is negative for the sleptons~\cite{randall}. 
In order to avoid this problem
we need to consider other positive soft contributions to the spectrum. 
This can arise in a number of ways, 
but any of the solutions will spoil the most attractive feature of anomaly
mediation, {\it i.e.} the RG invariance of the soft terms
and the consequent ultraviolet
insensitivity. This is, in our opinion, the most disappointing aspect
of these scenarios. Nevertheless, there are various options to cure this
problem without reintroducing the flavor problem. An example is the
inclusion of contributions from fields propagating in the bulk space
between the two branes~\cite{randall}. Another interesting 
possibility is a combination of gauge- and anomaly-mediated contributions,
discussed in ref.~\cite{pr}. 

The necessary cure for the slepton masses may completely upset also the
mass relations for the other particles (as in the case of the model
of ref.~\cite{pr}). 
However, 
here we will simply
parametrize the new positive contributions to the scalar
squared masses with a common mass parameter $m_0$,
assuming that the extra terms do not
reintroduce the supersymmetric flavor problem. 
We will see that many of the phenomenological
features of an anomaly-induced mass spectrum do
not crucially depend on the details of the contributions $m_0$.

\subsection{Defining a minimal model}
\bigskip

In the AMSB scenario, as discussed above,
the necessary mass parameters are the gravitino mass, 
$m_{3/2}$, and the common scalar mass
$m_0$, which is required to correct the negative mass-squared of the 
sleptons. The low-energy soft mass spectrum will be
\begin{eqnarray}
\label{spectroscopy}
     M_\lambda &=& {\beta_g\over g} m_{3/2}, \\
     m_{\tilde Q}^2&=&-{1\over 4}
      \left({\partial\gamma\over\partial g}\beta_g +
       {\partial\gamma\over\partial y}\beta_y\right) m_{3/2}^2 +m_0^2,
\label{budda} \\
\label{spectroscopy3}
     A_{y}&=&-{\beta_y\over y} m_{3/2}.
\end{eqnarray}
The expressions for the superpartner masses  of
the minimal particle content and soft parameters are given in the Appendix.
We will see that this soft-mass spectrum will give rise to distinctive 
features which differ from the usual gravity-mediated and gauge-mediated
scenarios.

Since our working framework is a theory with anomaly-mediated masses
and extra universal contributions to the scalar masses, we operationally
construct the full supersymmetric spectrum from four input parameters,
\beq
m_{3/2}, \, m_0, \, \tan\beta, \, {\rm sign}(\mu ).
\eeq
We treat the $\mu$ and $B_\mu$ masses as derived quantities that combine
with other terms in the scalar potential to reproduce correct electroweak
symmetry breaking (EWSB).  This procedure is done with the one-loop effective 
potential.  Also, we assume that Eq.~(\ref{budda}) is valid 
at the GUT scale. As previously discussed, the introduction of the
scalar mass $m_0$ breaks the RG invariance, and therefore we must define
a scale for the boundary condition Eq.~(\ref{budda}). 
Notice, however, that at the one-loop
level with Yukawa couplings neglected, the squark and slepton squared masses
are renormalized additively. Therefore, in this case, we do not need to
specify at which scale Eq.~(\ref{budda}) is valid. However this is not
true, for instance, for the stop and Higgs mass parameters.

We find that electroweak symmetry breaking can be accommodated 
with the above framework. Successful EWSB correlates
with values of $|\mu|$ typically
between 3 to 6 times the Wino mass as long as $m_0$ is not significantly
higher than the anomaly-mediated contributions to the squark masses.
Otherwise, $|\mu|$ can be larger.  The relative size of $\mu$ with
respect to $M_2$ becomes important when considering mass splitting
among the degenerate Wino triplet states.  This will be considered in  more
detail in the next section.

In Fig.~\ref{spectrum} we demonstrate
a subset of superpartner masses using a generically chosen set of
input parameters, $m_{3/2} =36\tev$, $\tan\beta =5$, and $\mu < 0$.
The choice of $m_{3/2}=36\tev$ determines the gaugino masses to
be $M_1=333\gev$,  $M_2=119\gev$, and $M_3 =850\gev$.
We vary $m_0$ to demonstrate its dependence in the scalar mass spectrum.
The squark masses are rather insensitive to values of
$m_0$ that raise the slepton masses above their anomaly-mediated
tachyonic values.    The sleptons, $\tilde e_L$ and $\tilde e_R$, 
are nearly equal in mass.  The extraordinary degeneracy of these slepton
masses will be expounded upon in the following section.
  
In Fig.~\ref{spectrum} we also plot the lightest physical Higgs boson mass,
$m_h$.
This is roughly constant over the range of $m_0$, since this eigenvalue
admits only logarithmic sensitivity to supersymmetry breaking scales.  
Requiring $M_2>90\gev$ and assuming $\tan\beta > 1.8$ (for perturbative
unification at the GUT scale), we find a lower bound on the lightest
scalar Higgs boson mass of $70\gev$.  The lower bound exceeds $100\gev$
for $\tan\beta > 5$. The upper bound on the Higgs boson mass, assuming
$M_2<500\gev$ and $m_0<m_{\tilde q}$ is $125\gev$.  However, 
the squark masses are
above $3\tev$ when the bound is saturated.  Since such high squark masses
are not welcome in the loop-corrected Higgs potential, the Higgs mass
is expected to be lighter than $125\gev$ in AMSB.
On the other hand, the pseudoscalar Higgs mass, $m_A$, depends linearly 
on the supersymmetry breaking
scale, and therefore increases with $m_0$ as shown in the figure.
In the next section, we study a few of the unique features of the AMSB
spectrum, and how it impacts search capabilities at high-energy colliders.
\jfig{spectrum}{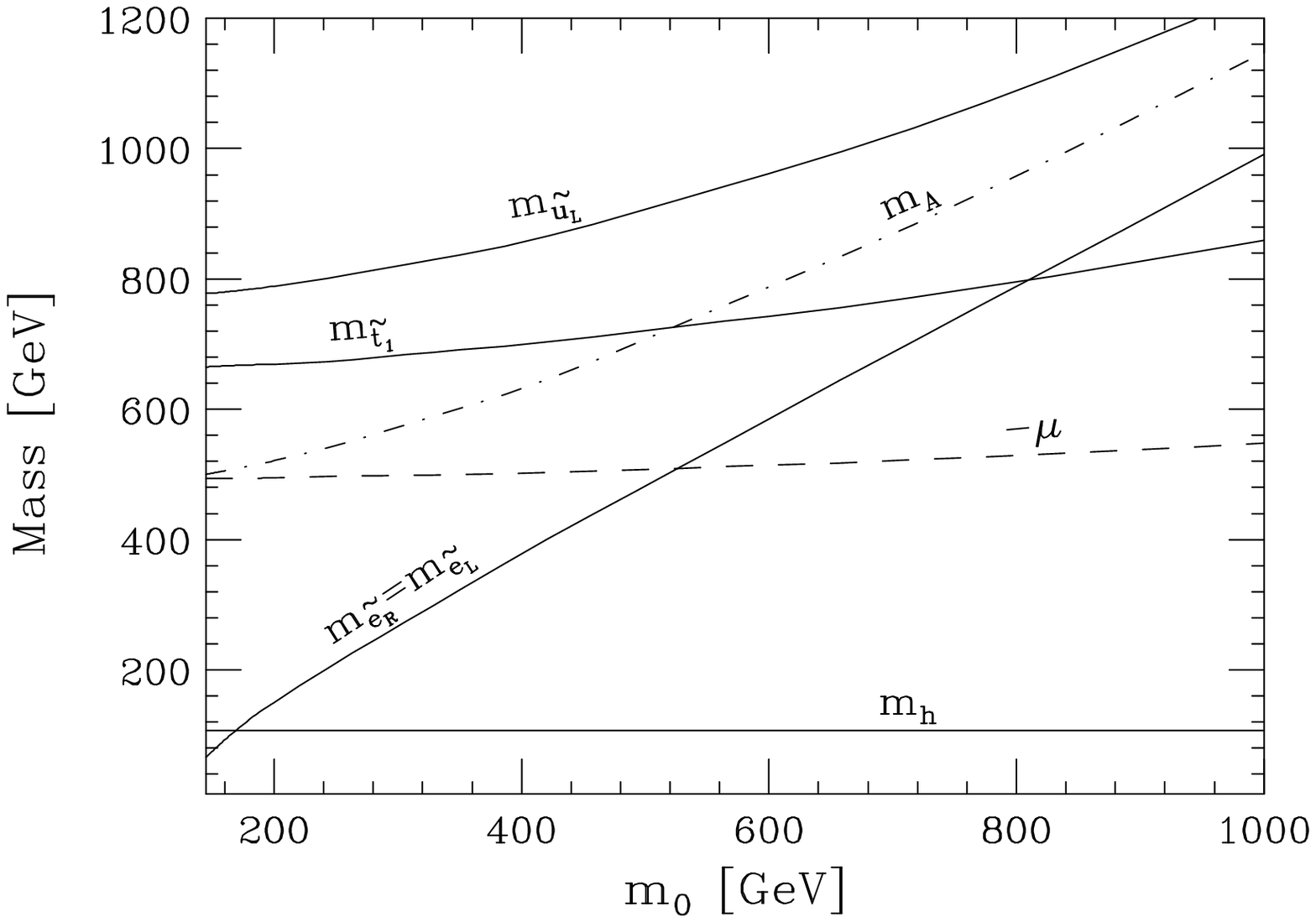}{ 
Masses of several states in the supersymmetric
spectrum as a function of $m_0$ with $m_{3/2} =36\tev$, 
$\tan\beta =5$, and $\mu < 0$.  
The gaugino masses for this choice are $M_1=333\gev$,  $M_2=119\gev$, 
and $M_3 =850\gev$.}

\section{Phenomenology}
\medskip

A unique feature of anomaly-mediated supersymmetry is the 
gaugino mass hierarchy.  
To compute the gaugino masses we include next-to-leading corrections coming 
from $\alpha_s$ and $\alpha_t \equiv y_t^2/4\pi$ two-loop contributions
to the beta-functions and weak threshold corrections enhanced by a
logarithm. In this approximation, we find 
\bea
M_1^{NLO}&=& M_1(Q)\left\{ 1+\frac{\alpha}{8\pi \cos^2\theta_W}
\left[ -21 \ln \frac{Q^2}{M_1^2} +11 \ln \frac{m_{\tilde q}^2}{M_1^2}
+9\ln \frac{m_{\tilde \ell}^2}{M_1^2}
\right. \right. \nonumber \\
&+& \left. \left. \ln \frac{\mu^2}{M_1^2}  
+\frac{2\mu}{M_1}\sin 2\beta \frac{m_A^2}{\mu^2-m_A^2}
\ln \frac{\mu^2}{m_A^2}\right] +\frac{2\alpha_s}{3\pi} -\frac{13\alpha_t}{66\pi}
\right\} \\
M_1(Q)&=&\frac{11\alpha (Q)}{4\pi \cos^2\theta_W} m_{3/2}
\eea
\bea
M_2^{NLO}&=& M_2(Q)\left\{ 1+\frac{\alpha}{8\pi \sin^2\theta_W}
\left[ -13 \ln \frac{Q^2}{M_2^2} +9 \ln \frac{m_{\tilde q}^2}{M_2^2}
+3\ln \frac{m_{\tilde \ell}^2}{M_2^2}
\right. \right. \nonumber \\
&+& \left. \left. \ln \frac{\mu^2}{M_2^2} 
 +\frac{2\mu}{M_2}\sin 2\beta \frac{m_A^2}{\mu^2-m_A^2}
\ln \frac{\mu^2}{m_A^2}\right] +\frac{6\alpha_s}{\pi} -\frac{3\alpha_t}{2\pi}
\right\} \\
M_2(Q)&=&\frac{\alpha (Q)}{4\pi \sin^2\theta_W} m_{3/2}
\eea
\bea
M_3^{NLO}&=& M_3(Q)\left\{ 1+\frac{3\alpha_s}{4\pi}
\left[ \ln \frac{Q^2}{M_3^2} + F\left( \frac{m_{\tilde q}^2}{M_3^2}\right)
-\frac{14}{9}\right] +\frac{\alpha_t}{3\pi} \right\} \\
F(x)&=& 1+2x+2x(2-x)\ln x +2(1-x)^2\ln |1-x| \\
M_3(Q)&=&-\frac{3\alpha_s (Q)}{4\pi} m_{3/2} .
\eea
The higgsino corrections to $M_1$ and $M_2$ are proportional to
$\mu / M_{1,2}$ and can become very important in models with large $\mu$,
as discussed in ref.~\cite{giudice}. The NLO corrections are significant,
especially for $M_2$ where the $6\alpha_s/\pi$ contribution changes the
Wino mass by more than $20\%$.

The mass ratios of the gauginos $M_1$:$M_2$:$|M_3|$
are approximately $3.3:1:8.8$ at leading order.  At NLO, these ratios
are changed to $2.8:1:7.1$.
This implies that a nearly degenerate
triplet of Winos ($\tilde W^\pm$, $\tilde W^0$) are the lightest
gauginos.  We shall see below that the neutral $\tilde W^0$ is
the lightest in the triplet, and is a candidate lightest supersymmetric
partner (LSP).  In an R-parity conserving theory the $\tilde W^0$ 
is stable and escapes detection at a high-energy collider.  Therefore,
visible particles produced in association with the $\tilde W^0$ states will
be required to uncover evidence of supersymmetry.  

It is also possible that the LSP is a sneutrino.  This would be the case
if the additional contributions to the scalar masses were large enough to
generate a positive mass-squared for the sleptons but still 
smaller than the Wino
mass. In this case, 
Wino decays would generally produce leptons and sneutrinos in
the final state. We will consider this possibility in some detail
in sect.~\ref{seclep}.

\subsection{Mass splitting among Winos}

The first step in considering light Wino states is to calculate the
mass splitting between the charged and neutral states.  For the 
moment we shall ignore loop corrections and describe the tree-level splitting
that develops for light Wino states.  Upon integrating out the heavy Bino
and Higgsino states, we are left with an effective theory with several 
operators
that could shift the mass of the remaining chargino and neutralino states
to be different than $M_2$.  
Operators of the form ${\cal O}=M_{ab}\tilde W^a\tilde W^b$
will generate mass splittings for the Winos only if $M_{ab}$ transforms
non-trivially under $SU(2)$.  
Because of the symmetry property of the Majorana mass term, $M_{ab}$ must
have isospin 2, and
the lowest-dimensional operator which 
generates a mass splitting is 
\beq
{\cal O}=\frac{1}{\Lambda^3}(H^\dagger \tau^a H)\, (H^\dagger \tau^b H)\,
    \tilde W^a\tilde W^b ,
\label{spin}
\eeq
where $\Lambda\sim M_1,\mu$ and $H$ denote the Higgs doublets.  Therefore, 
we see from the above that all mass splittings
at tree-level must occur with $m_W^4/\Lambda^3$ suppression.  

A more detailed formula for the tree-level mass splitting\footnote{To generate
a Wino mass splitting, it is also necessary to break the global custodial
$SU(2)_V$ defined such that the matrix $\Phi=\pmatrix{H^0_d &  H^+_u\cr
H^-_d & H^0_u}$, constructed from the two Higgs doublets, transforms
as $\Phi \to V \Phi V^\dagger$ with $V$ unitary. The $\mu$ term is invariant,
since it can be written as $\mu \, {\rm det} \Phi$. The symmetry is preserved
by electroweak breaking, as long as $\tan \beta =1$, but it is broken
by hypercharge effects. Therefore 
Eq.~(\ref{cazz}) has to vanish in the limit $\tan{\beta}\to 1$ and 
$\tan \theta_W \to 0$.}
with $|\mu|\gg M_1,M_2,m_W$ is\footnote{Our sign convention for 
$\mu$ is set by $W=\mu(H_u^0H_d^0-H_u^+H_d^-)$.}
\bea
m_{\tilde \chi^\pm_1}-m_{\tilde \chi^0_1} & = &
 \frac{m_W^4\sin^22\beta}{(M_1-M_2)\mu^2}
   \tan^2\theta_W
 + 2\frac{m_W^4 M_2 \sin 2\beta}{(M_1-M_2)\mu^3  }
     \tan^2\theta_W  \nonumber \\
 & &\quad + \frac{m_W^6\sin^3 2\beta}{(M_1-M_2)^2\mu^3}
        \tan^2\theta_W (\tan^2\theta_W -1) + {\cal O}(\frac{1}{\mu^4}).
\label{cazz}
\eea
When this formula is valid and $\mu$ is determined by the electroweak-breaking
condition, the mass splitting is negligible compared to
the charged pion mass -- an important mass scale for the phenomenology
of Wino decays.  
In our numerical analysis, we will always calculate
the chargino and neutralino mass splittings from the exact formula and
not from the expansion in Eq.~(\ref{cazz}), given here only for
illustrative purposes. Notice also that, in the 
large $\tan\beta$ limit,  the Wino mass difference becomes
\beq
    m_{\tilde \chi^\pm_1}-m_{\tilde \chi^0_1} = 
  \frac{M_2 m_W^4}{2\mu^4}\left(1+\frac{2M_2\tan^2\theta_W}{M_1-M_2}\right)
  + {\cal O}(\frac{1}{\mu^6}),~~~~{\rm for}~\tan\beta \to \infty .
\eeq
In this limit the mass difference has a further suppression
factor, $M_2/\mu$ because the necessary chiral flip cannot originate from
the Higgsino mass.

The dominant 
contribution to the Wino mass splitting
does not come from the tree-level result described above, but 
rather due  to one-loop corrections in the 
chargino and neutralino mass matrices.
We have done a full numerical calculation of the one-loop corrected
chargino and neutralino mass matrices using the formulae
of ref.~\cite{pierce}.  For the anomaly-mediated spectrum, with only positive
mass-squared additional contributions to all the scalar masses, 
we find that 
the gauge-boson loop corrections dominate the mass splitting. This is because
in the typical anomaly-induced mass spectrum the squark masses are heavy
and the $\mu$ parameter is large. Consequently, following the argument
that led us to Eq.~(\ref{spin}), we infer that their contribution to the
Wino mass splitting is suppressed by $M_W^4/\Lambda^3$. On the other hand,
the effect of gauge-boson loops cannot be described by local operators.
Isolating this contribution in the limit of large $\mu$,  we find (see also
refs.~\cite{cheng,feng})
\beq
\label{gaugediff}
    \Delta_\chi\equiv m_{\tilde\chi_1^\pm} - m_{\tilde\chi_1^0} 
     ={\alpha M_2\over\pi\sin^2\theta_W}
    \left[f(m_W^2/M_2^2)-\cos^2\theta_W
    f(m_Z^2/M_2^2)\right]
\eeq
where 
\beq
    f(x)\equiv-{x\over 4}+{x^2\over 8}\ln x+{1\over 2}(1+{x\over 2})
    \sqrt{4x-x^2}
    \left[{\rm arctan}{2-x\over\sqrt{4x-x^2}}-{\rm arctan}{x\over\sqrt{4x-x^2}}
    \right].
\eeq
In the limit that $M_2\rightarrow\infty$, the expression in 
Eq.~(\ref{gaugediff}) 
simply becomes
\beq
\label{loop split}
   \Delta_\chi= \frac{\alpha\, m_W}{2(1+\cos\theta_W)} 
   \left[1-{3\over 8\cos\theta_W}
   {m_W^2\over M_2^2}+{\cal O}\left({m_W^3\over M_2^3}\right)\right],
\eeq
which has the asymptotic limit $\Delta_\chi=\alpha\, m_W/[2 (1+\cos\theta_W)]
\simeq 165$ MeV.

It may appear odd that the mass splitting should asymptote to a constant 
value as
$M_2$ gets arbitrarily massive.  This behavior can be understood in momentum
space as an infrared mismatch between the self-energies of
$\tilde W^+$ and $\tilde W^0$ regulated by $m_W$.
Or, equivalently, since 
$SU(2)$ is a good theory for short distances $r\ll m_W^{-1}$, we can
calculate the Coulomb energy of the charged state for large distances
$r\gsim m_W^{-1}$ (infrared region) to obtain a mass splitting
of approximately $\alpha m_W$.  The exact prefactors are given
in Eq.~(\ref{loop split}).

In Figs.~\ref{split_tb2} and~\ref{split_tb40} 
we show the total calculated mass splitting 
as a function of $M_2$ for $\tan\beta =2$ and~10.  In our numerical
calculation, we include the full one-loop result and we do not use the
approximate expressions given in Eq.~(\ref{gaugediff}).
The solid curves, from top to bottom, represent
$\mu=2M_2$, $\mu=3M_2$, $\mu=5M_2$, and $\mu=\infty$.  The dashed curves
are the same except that $\mu <0$.
The dot-dashed curve is
the charged pion mass $m_{\pi^\pm}$.  As $\tan\beta$ increases the
sign of $\mu$ becomes less and less relevant in the calculation of
the mass splitting.  When $\tan\beta =40$ the solid and dashed curves
are irresolvable.
\jfig{split_tb2}{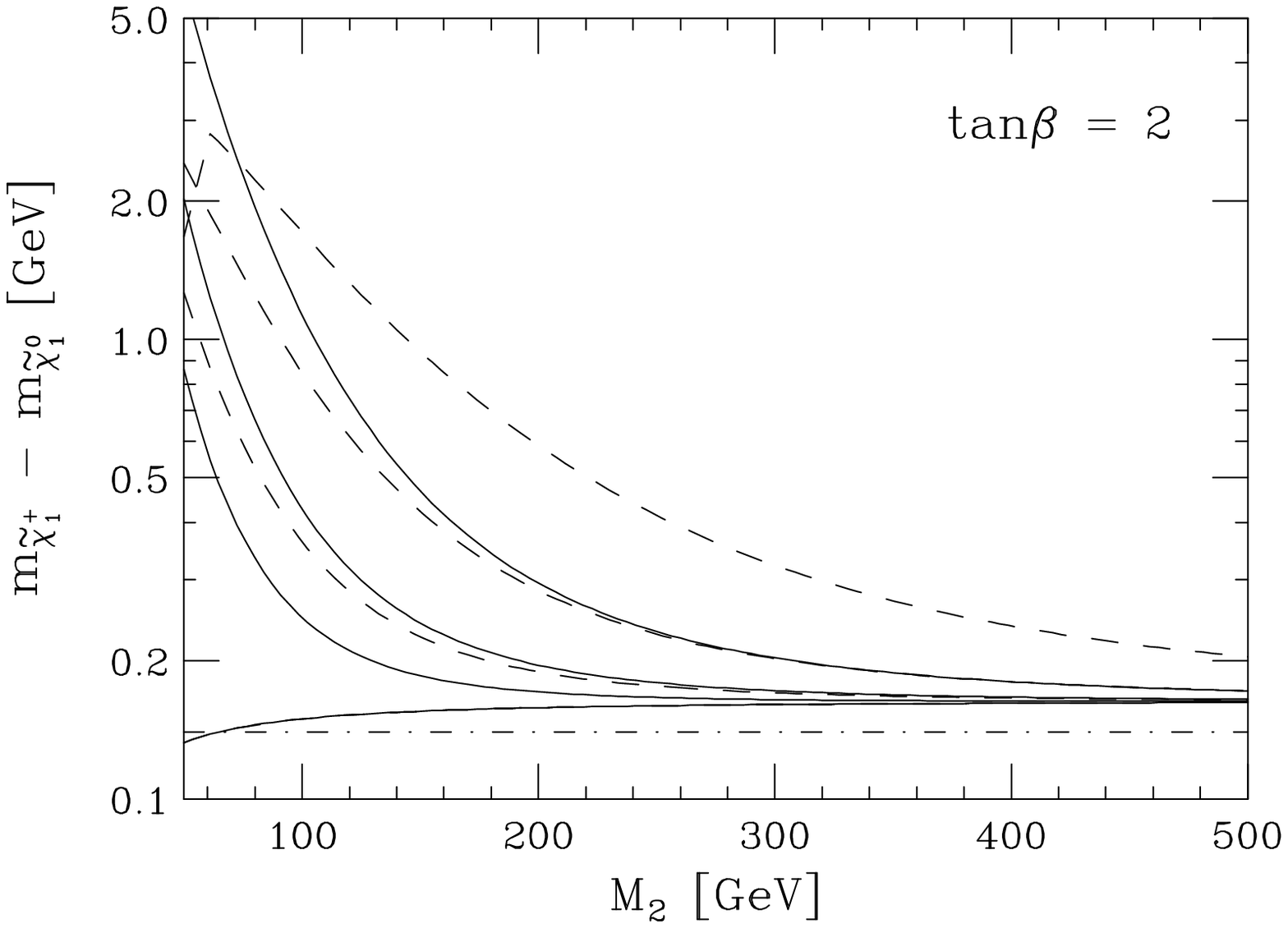}{\protect 
The mass splitting as a function of $M_2$ for
$\tan\beta =2$.  The solid curves, from
top to bottom, represent
$\mu=2M_2$, $\mu=3M_2$, $\mu=5M_2$, and $\mu=\infty$.  The dashed curves
are the same except for the opposite sign of $\mu$.  The dot-dashed curve is
the charged pion mass $m_{\pi^\pm}$.}
\jfig{split_tb40}{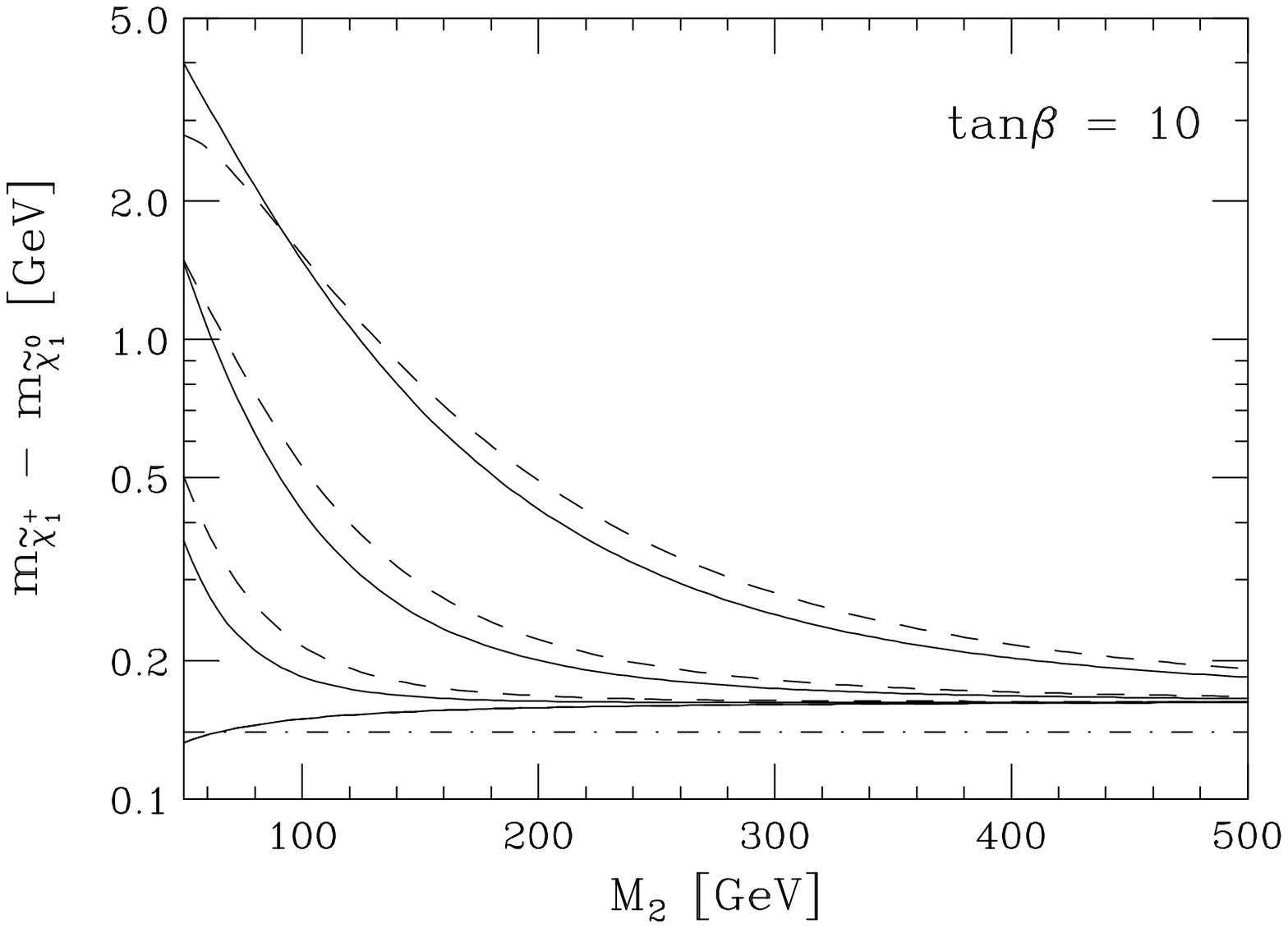}{\protect 
The mass splitting as a function of $M_2$
for $\tan\beta =10$.  The solid curves, from
top to bottom, represent
$\mu=2M_2$, $\mu=3M_2$, $\mu=5M_2$, and $\mu=\infty$.  The dashed curves
are the same except for the opposite sign of $\mu$.  The dot-dashed curve is
the charged pion mass $m_{\pi^\pm}$.}

In an anomaly-mediated spectrum with radiative electroweak symmetry 
breaking, the
typical relation between $M_2$ and $\mu$ is $3\lsim |\mu|/M_2\lsim 6$.
This is true as long
as the squark masses are not increased significantly beyond their
anomaly-mediated baseline values from the universal mass 
contributions that lift the
slepton mass-squared to  positive values.  This is also acceptable from naive
fine-tuning arguments.  Larger values of $\mu$ lead to an unnatural
Higgs potential.  For $|\mu|/M_2= 5$ we
find from Figs.~\ref{split_tb2} and~\ref{split_tb40} 
that the mass splitting is significantly 
above $m_{\pi^\pm}$ such that $\tilde W^\pm \to \tilde W^0\pi^\pm$ is
kinematically allowed and is the dominant decay mode.
This remark is also
true even for extraordinarily large values of $\mu$ as long as $M_2\gsim
80\gev$.  

\subsection{Finding supersymmetry with dileptons}

The precise calculation of the mass splitting is crucial since in
ref.~\cite{gunion} it was demonstrated that if 
$m_{\pi^\pm}\lsim m_{\tilde \chi^\pm_1}-m_{\tilde \chi^0_1} \lsim 1\gev$
then the $\tilde W^\pm$ will decay too fast to use a quasi-stable charged
particle analysis, with dedicated triggers.  
However, the decays are not prompt, and so analyses
of events triggered by other means could see a stiff charged particle 
track that
subsequently terminates in the vertex detector. The difficulty is triggering
the event.

One way to trigger such events is to produce the Winos in associated production
with a standard model particle, such as a gluon at hadron colliders or 
a photon at $e^+e^-$ colliders. Triggering on high-$p_T$
monojets or high-energy photons at these colliders  then may be an 
effective way
to trigger the events and save them for future 
analysis~\cite{gunion,thomas,feng}. 
At the analysis
stage a kink in the vertex detector, or a terminating stiff track,
would then indicate a non-SM underlying
process.

Here we pursue another direction for discovery.  We can
utilize production and subsequent decays of other SUSY particles as a way
to trigger on the events and learn more about the theory.  For example, if
sleptons or squarks are produced in a hadronic collision, they will 
cascade decay
to high-$p_T$ SM particles and charged and/or neutral Winos.  The SM particles
can be used for the trigger, and the cascade decays can be used to 
learn something
about the spectroscopy of the theory.  

Our example process is left-handed slepton and sneutrino production at the
Tevatron which
cascades into $l^\pm l^\mp+X_{D}$, where $X_D$ 
is a displaced vertex from
one
or two $\tilde W^\pm \to \tilde W^0\pi^\pm$ decays.  These displaced vertices
are heavy charged particle tracks which stop in the vertex detector and produce
very soft pions that may or may not be detectable. 
The dilepton events are produced through
\bea
p\bar p \to \tilde \nu_L\tilde l^{\pm}_L & \to & 
        l^\pm l^\mp \tilde W^\pm\tilde W^0
                   \to l^\pm l^\mp +X_D+\mEt \\
p\bar p \to \tilde \nu_L\tilde \nu_L & \to & 
             l^\pm l^\mp \tilde W^\pm\tilde W^\mp
                   \to l^\pm l^\mp +X_D+\mEt .
\eea
In Fig.~\ref{likesign} we plot the total cross-section of such events for one
flavor ($\mu^\pm\mu^\mp+X_D$) 
at the $\sqrt{s}=2\tev$ Tevatron.  We
require
the pseudo-rapidity to be within $|\eta|<2$ for both leptons, and we
require the leading lepton to have $p_T>10\gev$, and the next lepton to have
$p_T>5\gev$.  
The total rate presented in Fig.~\ref{likesign} is calculated
at leading order.  
We have included a total of 50\% suppression of the naive LO result
from jet veto and lepton identification efficiency.

Our conclusion based on Fig.~\ref{likesign} is that left-handed sleptons with
mass 
less than $200\gev$ would be discovered at the Tevatron if
$M_2\lsim m_{\tilde \nu_L}-10\gev$ (for leptons to have high enough $p_T$ for
triggering) and if the Tevatron reaches at least $30\xfb^{-1}$ integrated
luminosity. This result is based on the requirement that more than 10
$l^\pm l^\mp+X_D$ events will occur for each flavor. 
We conservatively
choose a 10 event requirement in order to ensure that our
mass reach conclusion will remain valid if the dilepton identification
efficiency were to be  
as low as $50\%$ for the $p_T$ acceptance cuts given above.

Other modes such as $\mu^\pm +X_D$ are
possible in $\tilde \nu_L$ and $\tilde \mu_L$ production, and could confirm
and extend the mass reach capabilities of the dilepton mode.
\sfig{likesign}{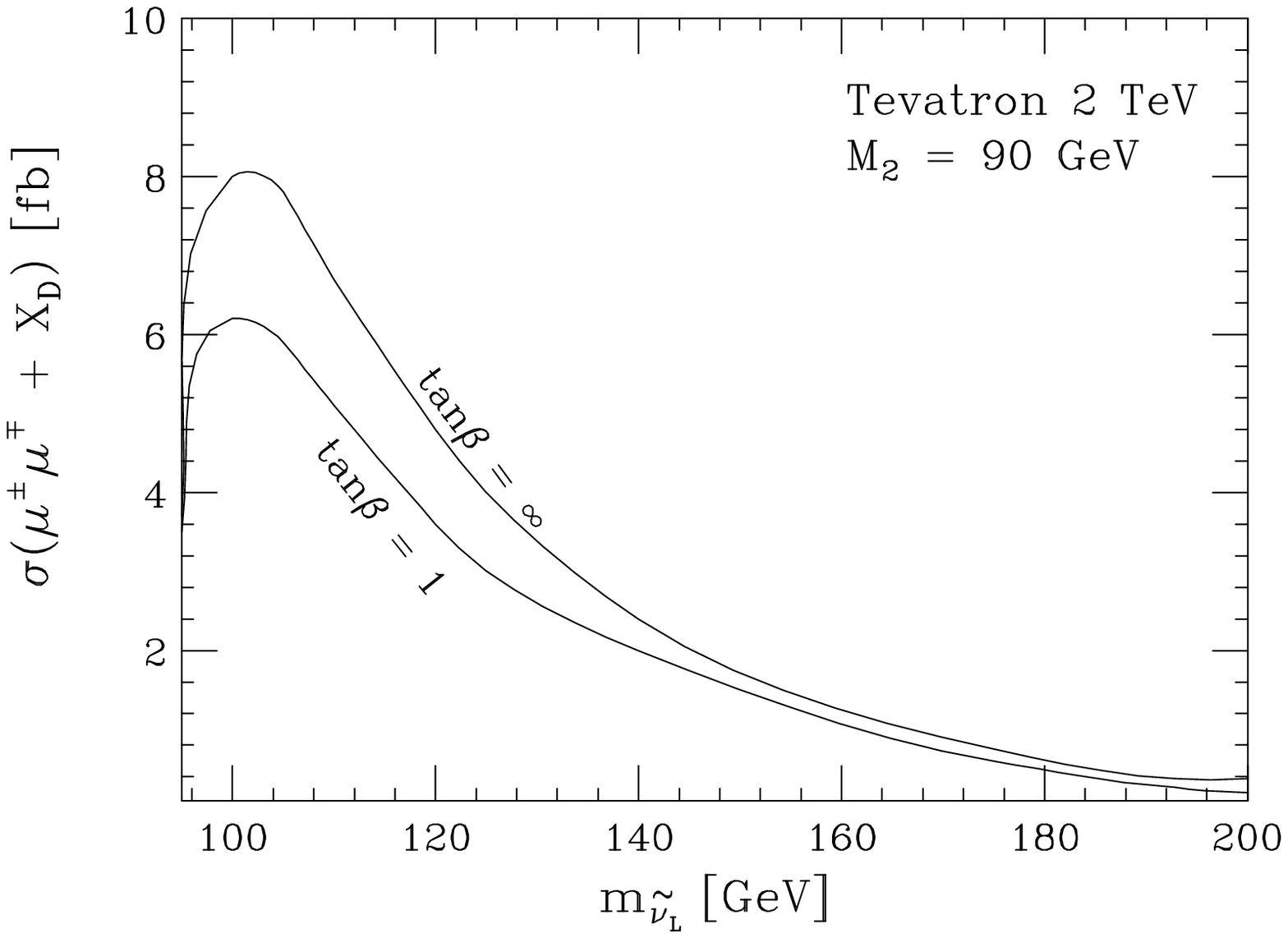}{Dilepton signal from 
left-handed smuon and sneutrino production at the Fermilab Tevatron
with $2\tev$ center of mass energy and $M_2=90\gev$.  
Acceptance cuts of the leptons are
described in the text.  The different curves are for $\tan\beta =1$,
which makes $m_{\tilde \mu_L}=m_{\tilde \nu_L}$, and for $\tan\beta = \infty$
which maximizes the hypercharge $D$-term splitting such that
$m_{\tilde \mu_L}^2=m_{\tilde \nu_L}^2+m_W^2$. 
With $30\xfb^{-1}$ the Tevatron will record more
than 10 such events for each lepton flavor if $m_{\tilde \nu_L}<200\gev$.}

The dilepton signal discussed above is special since it is
essentially background
free with the displaced vertices present.  The possibility 
of prompt Wino decays arises
if $\mu$ is sufficiently light to yield a chargino/neutralino mass
splitting above $1\gev$.  This would make the charged Wino decay promptly into
a neutral Wino plus other states too soft to admit into the event description.
Also, if the top squarks are reduced from additional negative
scalar mass sources, the
mass splitting could be greatly enhanced by loop corrections involving
third family sfermions and fermions.  In these
cases, special triggers or analyses based on decay kinks of the charged
Wino could not be relied upon, but  the dilepton signal would
remain useful.

\subsection{Degeneracy of sleptons}

Another striking feature of the anomaly-mediated model with additional
universal scalar terms is the near degeneracy of the left and right sleptons
of the first two generations.  The mass-squared splitting is 
somewhat insensitive to $m_0$,
\bea
\label{ersplit}
\Delta_{\tilde e} & = & m^2_{\tilde e_L}-m^2_{\tilde e_R} = 
 (11\tan^4\theta_W-1)\frac{3}{2}M_2^2
 + \left(-\frac{1}{2}+2\sin^2\theta_W\right) m^2_Z\cos 2\beta \nonumber \\ 
 & & \qquad\qquad\quad\quad\quad
      +\frac{1}{8\pi^2}\left( \frac{9}{5}g_1^2 M^2_1 - 3g_2^2 M_2^2\right)
       \ln \frac{m_{\tilde e_R}}{m_Z} \nonumber \\ 
 & \simeq & 0.037
   \left( -m_Z^2\cos 2\beta+M_2^2 \ln \frac{m_{\tilde e_R}}{m_Z} \right).
\eea
The first term is the tree level anomaly-induced splitting, the second
term is the hypercharge $D$-term splitting induced by electroweak
breaking, and the third
term is the one loop, leading log mass splitting induced by renormalizing
the masses to their own scale.
It is a numerical accident that the value of $\sin^2\theta_W$ 
is such 
that the $M_2^2$ coefficient in the first term of Eq.~(\ref{ersplit})
is nearly zero.  If,
\beq
\label{sin squared}
\sin^2\theta_W=\frac{1}{1+\sqrt{11}}=0.2317
\eeq
then the tree-level 
coefficient of $M_2^2$ would be identically zero.  The actual
value of $\sin^2\theta_W(m_Z)$ is $0.2312\pm 0.0003$~\cite{PDG} in
the $\overline {MS}$ scheme and is extraordinarily close to the value
in Eq.~(\ref{sin squared}) required to make the $M_2^2$ coefficient in the
tree-level mass splitting vanish.  

It is also a numerical accident that the hypercharge $D$-term 
coefficient is suppressed since $\sin^2\theta_W\simeq 1/4$.  
Although the coefficient is not
as spectacularly suppressed as the $m_{3/2}^2$ coefficient, it is
multiplied
by a fixed scale $m_Z^2$. Therefore, for a given value of $\tan\beta$
the mass squared difference remains constant regardless of how 
heavy the sleptons may be.  

The degeneracy of the slepton can be characterized by the fractional
difference, 
\beq
\frac{m_{\tilde e_L}-m_{\tilde e_R}}{m_{\tilde e_R}}
  = -1 +\sqrt{ 1+\frac{\Delta_{\tilde e}}{m^2_{\tilde e_R}}}
  ~\simeq \frac{1}{2}\frac{\Delta_{\tilde e}}{m^2_{\tilde e_R}} .
\eeq
In Fig.~\ref{diff} we plot contours of the relative mass splitting in the 
$M_2$-$m_{\tilde e_R}$ plane.  The mass splitting is less than
a few percent over most of parameter space.  It exceeds 5\% only
when $M_2>350\gev$.  However, the squark masses in this case
are over $2\tev$, which induces a considerable
fine-tuning in the one-loop Higgs potential.
Therefore, it is not expected that $M_2$ is so high, implying that
the slepton mass differences should be no more than a few percent
over the relevant parameter space.
\jfig{diff}{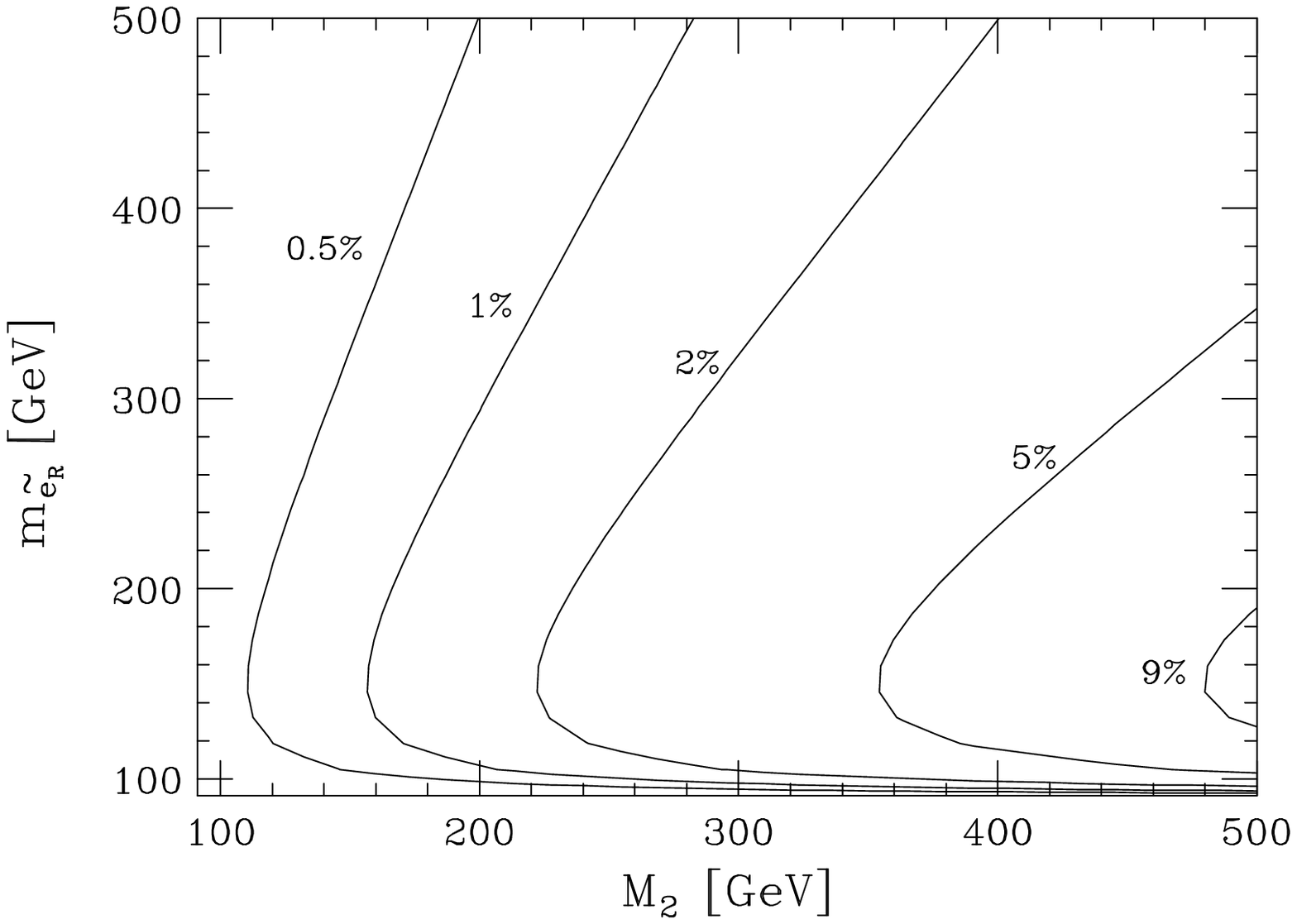}{Contours of $100\% \times 
(m_{\tilde e_L}-m_{\tilde e_R})/m_{\tilde e_R}$ 
in the $M_2$-$m_{\tilde e_R}$ mass plane with large
$\tan\beta$, which maximizes the mass splitting.}

The resolution of slepton masses from end-point lepton
distributions of the slepton decays
is approximately $2\%$ at an $e^+e^-$ or a muon 
collider~\cite{NLC report}.
Note that at a polarized linear collider it will not be difficult to
determine that both left and right sleptons are being produced
even if they are degenerate.  This can be accomplished most effectively
by comparing
total rate of sleptons production with the asymmetry of production
for polarized beams~\cite{NLC report}.
Degenerate sleptons are not expected in 
the usual supergravity or gauge-mediated scenarios, where 
$m_{\tilde e_R}$ is  generally lighter
than $m_{\tilde e_L}$.  An exception to this is in the
minimal supergravity model with $m_0\gg m_{1/2}$.  However,
given the current limits on $m_{1/2}$ from gaugino searches
at the Tevatron and LEP, degenerate sleptons will only
occur at very high mass.

The above discussion is based on the assumption that the additional
contributions to the slepton masses are universal. Since the 
anomaly-mediated mass differences is accidentally negligible, 
the degeneracy of the sleptons becomes a test of 
the additional mass contributions.  Other approaches
to the slepton tachyonic problem do not necessarily imply
degenerate slepton mass eigenstates~\cite{pr}.

\subsection{LEP2 Signals} \label{seclep}
\bigskip

At LEP2 many signatures are possible in the AMSB scenario.  The charged Winos
have a large production cross-section as long as they
are kinematically accessible and as long as the sneutrino $t$-channel
amplitude does not significantly interfere destructively with gauge
boson $s$-channel amplitudes.  Production of charginos at LEP has been
the topic of many studies in supersymmetry phenomenology at LEP~\cite{leprep}.
However, most of these studies have assumed that the LSP, the lightest
neutralino, is more than a few GeV below the lightest chargino mass.
In AMSB this is no longer the case.  We expect that the lightest neutralino
and charginos form a nearly degenerate $SU(2)$ triplet, as discussed
in previous sections.  

Since the chargino is only slightly above the neutralino in mass, the
process $e^+e^-\to \tilde W^+\tilde W^-$ will be accompanied by very
soft visible final states from decays such as $\tilde W^+\to \pi^+\tilde W^0$.
These soft final states cannot be triggered on, which has led 
others~\cite{gunion,thomas,feng}
to suggest triggering on initial state photon radiation and then
searching for soft, displaced tracks at the analysis level.  However,
there are
other potential signatures of supersymmetry when Winos are the lightest
gauginos.  To enumerate them we must consider the other states in the 
supersymmetric spectrum which may be produced in collisions or as
decay products of Wino production.  In the AMSB scenario, the
degenerate left and right sleptons and the sneutrino are the most
important states at LEP after the gauginos.  The ratio of their masses
to the Wino masses is unknown in our framework, but it is more natural
that they be somewhat light in order to keep the Higgs scalar potential
from being fine-tuned.  Therefore, considering phenomenological implications
of light sleptons at LEP is useful.

There are many permutations to the  
relative ordering of $M_2$, $m_{\tilde \nu_L}$, and 
$m_{\tilde e}\equiv m_{\tilde e_{L,R}}$.  Recall that the relationship
between $m_{\tilde e}$ and $m_{\tilde\nu_L}$ is
\beq
m^2_{\tilde e_L}=m^2_{\tilde\nu_L}-m^2_W\cos 2\beta.
\eeq
We can provide the
general phenomenological features using a graph in the $M_2$-$m_{\tilde\nu_L}$
plane for LEP2 running at $\sqrt{s}=200\gev$. 
The results of the present and future experimental analyses combining
the searches at LEP2 in different channels will be best presented as
exclusion or discovery regions in the $M_2$-$m_{\tilde\nu_L}$
plane.
The LEP1
limit of Wino masses is slightly above
$m_Z/2$, and the limit on sneutrino masses is slightly below
$m_Z/2$; therefore, we begin the axes of Fig.~\ref{m2msnu} at $m_Z/2$.  
The lines represent kinematic boundaries.  For example, the top dashed
line is where $m_{\tilde e}=\sqrt{s}/2$, and for all $m_{\tilde\nu_L}$
values above that line, $\tilde e\tilde e$ production is not possible.
The precise locations of the
dashed lines depend on the choice of $\tan\beta$, which we choose
to be $\tan\beta =3$ for this figure.
\bfig{m2msnu}{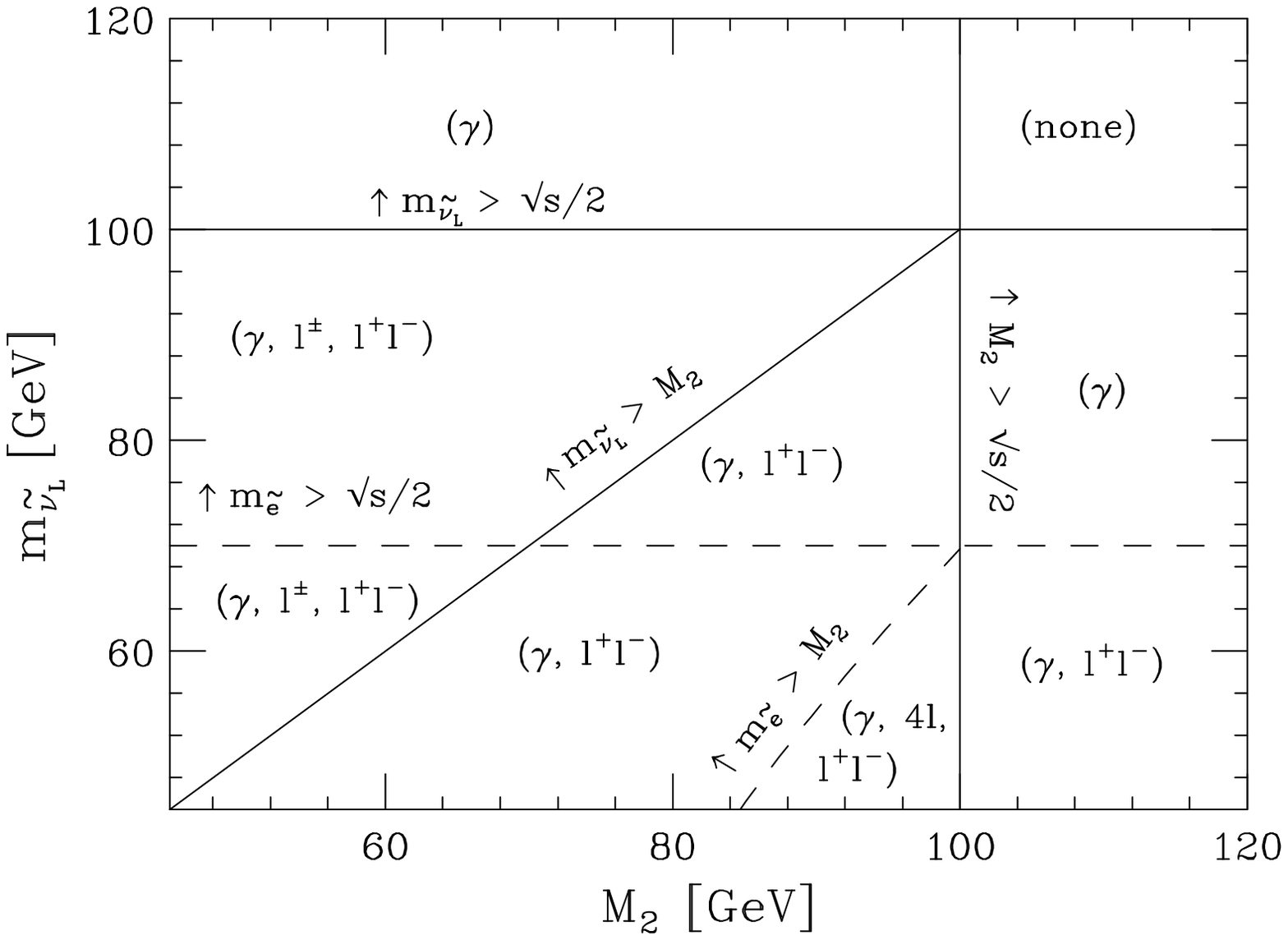}{Signatures of the AMSB scenario at LEP2.
Each blocked region has a unique mass hierarchy among $\tilde \nu_L$,
$\tilde e_L$ and $M_2$, and therefore leads to different signatures
which are contained within the parenthesis.   The meaning of ``$(\gamma )$'',
for example, is $e^+e^-\to \gamma \tilde W^+\tilde W^-\to \gamma +\mE$.
We assume, as is justified by EWSB analysis,
that $M_2$ is very close to the mass of the nearly degenerate lightest
charginos and neutralino.  The mass splitting between $m_{\tilde \nu_L}$
and $m_{\tilde e}$ is due to a hypercharge $D$ term, and its value
is calculated with $\tan\beta =3$ for this figure.}

Each region in the figure 
constitutes a different ordering of the mass eigenstates,
and will produce a different set of useful observables with which
to probe the theory.  These observables are given inside the
parentheses. For example, in the small triangular region 
surrounding the point $(M_2,m_{\tilde\nu_L})=(95\gev ,60\gev )$ one
can search for $\gamma +\mE$, $l^+l^-+\mE$, and up to four electrons
plus missing energy.  Specific processes which lead to these signatures 
include,
\bea
e^+e^- & \to & \gamma \tilde \nu_L\tilde\nu_L \to \gamma +\mE \\
e^+e^- & \to & (\tilde W^+ \tilde W^-~{\rm or}~
       \tilde l^-\tilde l^+)\to l^+l^-+\mE \\
e^+e^- & \to & \tilde W^0 \tilde W^0\to \tilde l^+ l^-\tilde l'^+ l'^-\to
                  l^+l^- l'^+l'^- +\mE. 
\eea
Some of the leptons may be softer than others because of reduced 
phase space in a decay of a massive sparticle into a lepton and a sparticle
with mass near its parent.  Near the boundaries of the curves, it is often
the case that some leptons are not energetic enough,
and care must be taken in the analysis to identify them.

Finally, the stau sleptons may be lighter than the other sleptons, leading 
to more $\tau$ lepton final states than other leptons.  Although efficiency
in identifying $\tau$ leptons is smaller than the others, it is possible
at large $\tan\beta$ to have large mixing among the $\tilde \tau_L$
and $\tilde \tau_R$ sleptons to produce a mass eigenstate accessible to LEP
whereas the other sleptons are not.

\section{Gravitino cosmology} 
\medskip

A distinctive feature of the AMSB scenario is that the gravitino 
is much heavier than the supersymmetric partners of ordinary particles.
The reason for this is that the AMSB masses
are suppressed by a loop factor relative to
the gravitino mass $m_{3/2}$. 
In particular, one finds that the gravitino--Wino mass ratio is
$m_{3/2}/M_2 \simeq 300$.

A large gravitino mass is cosmologically advantageous for solving the 
gravitino problem \cite{weinberg}. This problem occurs when
gravitino decay products disrupt the light element abundance during
nucleosynthesis. Even a period of inflation is not sufficient to solve
this problem since gravitinos are thermally produced 
during the reheating phase of the universe~\cite{ellis}. Thus, in order 
for the gravitino decay products to be harmless during
nucleosynthesis one either requires that the gravitino decays before 
affecting nucleosynthesis or that the reheating temperature of the universe 
be bounded from above.

The gravitino number density in units of the photon number density after
the inflationary epoch is~\cite{moroi}
\beq
\frac{n_{3/2}}{n_\gamma}(T)= 2\times 10^{-9} g_*(T)~ T_{13}( 1-0.03
\ln T_{13}) .
\eeq
Here $T_{13}$ is the reheating temperature after inflation
in units of $10^{13}\gev$ 
({\it i.e.}, $T_{13}\equiv T_R/10^{13}$ GeV), 
and $g_*(T)$ counts the massless degrees of freedom at the
temperature $T$ including
a factor of $7/8$ for fermions and the dilution factor for frozen-out species.

The gravitino decay width is
\beq
    \Gamma_{3/2} = {1\over 4}\left(N_g+{N_m\over 12}\right)
    {m_{3/2}^3\over M_P^2} \simeq 5.1
\left({m_{3/2}\over 50~{\rm TeV}}\right)^3 {\rm sec}^{-1}.
\eeq
Here $M_P=1.2\times 10^{19}$ GeV, 
$N_g$ and $N_m$ are the number of gauge and matter decay channels,
and 
we have summed over all the SM particle content 
($N_g$=12, $N_m$=49).

Immediately after the gravitino decays, the temperature of the universe is
given by
\beq
T_D=\left( \frac{45\, \Gamma_{3/2}^2M_P^2}{4\pi^3g_*(T_D)}\right)^{1/4}=
2.7 \left( \frac{10.75}{g_*(T_D)}\right)^{1/4}
\left({m_{3/2}\over 50~{\rm TeV}}\right)^{3/2} {\rm MeV}.
\eeq
Therefore for $m_{3/2}\lsim 60$ TeV a detailed analysis of effects
of gravitino decay during nucleosynthesis is necessary.

The particular gravitino decay products that cause the 
main interference during nucleosynthesis at early times ($\sim 1$ second)
are hadronic 
showers~\cite{reno}. Photodissociations are not relevant at early stages
of the nucleosynthesis epoch 
since the destructive photon-nucleus interactions are much less 
probable than photon-photon interactions.
The overall effect of hadronic decay products is to convert protons 
into neutrons, and consequently the $^4$He abundance is increased since
the additional neutrons that are produced are synthesised into $^4$He.
Thus, using the observational upper limit on
the primordial helium abundance $Y(^4{\rm He})<0.25$, we obtain an
upper bound on the reheat temperature~\cite{sarkar} which is 
depicted in Fig.~\ref{reheat}.
\jfig{reheat}{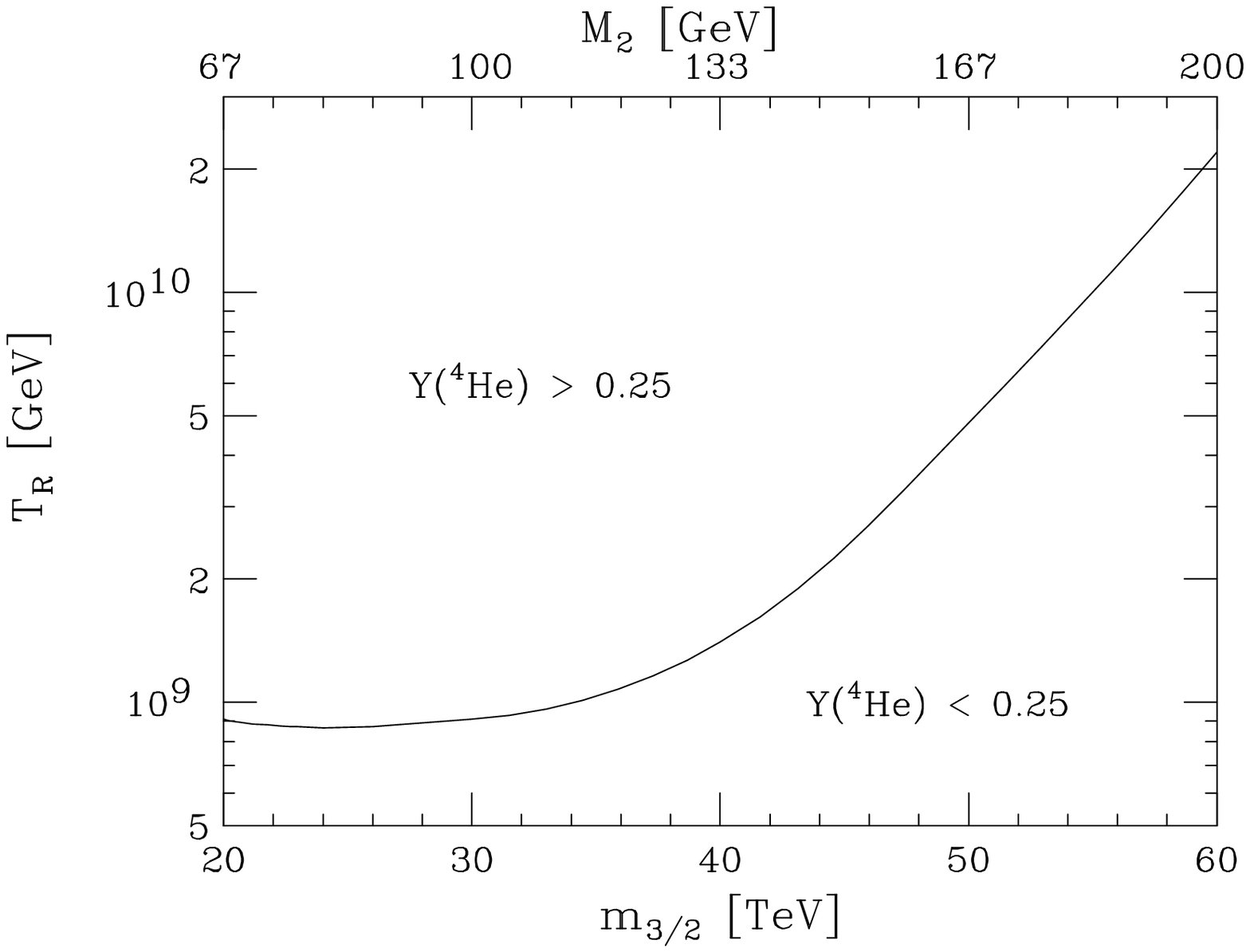}{The upper bound on the reheat temperature
as a function of the gravitino mass $m_{3/2}$ and the corresponding 
Wino mass $M_2$.  The observational limit upper limit on the 
primordial helium abundance is $Y(^4{\rm He})<0.25$.  Therefore,
the region above the line is excluded, and an upper bound on
$T_R$ results.}
For $m_{3/2}\lsim 40$ TeV, the typical bound on the reheat 
temperature is $T_R\sim 10^{9}$ GeV.  This
is typically less
constraining 
than in the usual gravity-mediated 
scenarios with weak-scale gravitino mass~\cite{sarkar}.

As the gravitino mass 
increases, the upper bound on the reheat temperature becomes less significant,
and completely evaporates for $m_{3/2}\gsim 60$ TeV,
since the gravitinos decay well before the start of nucleosynthesis.
For $m_{3/2}\gsim 60$ TeV, we may be concerned that the entropy
produced by the gravitino decay excessively dilutes the baryon--to--photon
ratio obtained by a primordial baryogenesis mechanism. However, this is
never the case. Actually  for
\beq
{m_{3/2}\over 50~{\rm TeV}} >8\times 10^{-4} \left(\frac{g_*(T_D)}
  {10.75}\right)^{1/2} 
   T_{13}^2( 1-0.03\ln T_{13})^2 ,
\eeq
the gravitino decays before dominating the universe and the entropy release
is not dangerous. We have checked that, even when the gravitino 
matter-dominates
the universe, the entropy production is not problematic. Therefore we conclude
that when $m_{3/2}\gsim 60$ TeV, there is no upper bound
on the reheat temperature arising from nucleosynthesis.

However, bounds on the reheating temperature 
when $m_{3/2}\gsim 60\gev$ do come from
considerations of the Wino energy density.
The Wino thermal relic abundance $\Omega_{LSP}^{TH}$ 
does not play a significant cosmological
role. Indeed Wino annihilations into gauge bosons in the early
Universe are very efficient and lead to~\cite{giudice}
\beq
\Omega_{LSP}^{TH} h^2 \simeq 5\times 10^{-4} \left( \frac{M_2}{100~{\rm GeV}}
\right)^2 .
\eeq
On the other hand, a non-thermal production of LSPs is generated by the 
gravitino decay~\cite{frie}. Since this decay occurs below the Wino freeze-out
temperature, the LSP abundance is easily determined by assuming that 
each decaying gravitino produces a single Wino. The LSP is relativistic
at decay time and becomes non-relativistic at a typical temperature
$T\sim T_D M_2/m_{3/2}$, after red-shifting. The predicted relic abundance
is
\beq
 \Omega_{LSP}^{G} h^2\simeq 30 \left(\frac{M_2}{100\, {\rm GeV}}\right)
    T_{13}(1-0.03\ln T_{13}).
\eeq
The requirement for not
overclosing the universe 
leads to a bound on the reheat temperature $T_R\lsim
10^{11}\gev$. Of course, if 
R-parity is not conserved then this bound from LSP relic abundance is
no longer relevant. 
In this case, for gravitino masses $m_{3/2}\gsim 60$ TeV,
one can then contemplate using 
leptogenesis or even GUT baryogenesis mechanisms 
to generate the baryon asymmetry of the 
universe, consistently with the AMSB gravitino cosmology.

On the other hand, when $T_R\simeq 10^{10}$--$10^{11}\gev$ 
the relic abundance of
LSPs from gravitino decays is near critical density, providing a natural
source of dark matter.  

Finally, we comment on another positive aspect of the heavy gravitino
cosmology. 
We can  avoid the cosmological Polonyi problem that arises in the 
usual gravity-mediated scenario when the gauge singlet Polonyi field
acquires a Planck scale vacuum expectation value 
but decays relatively late. In the AMSB 
scenario there is simply no need for the Polonyi field since the 
gaugino masses arise from a quantum anomaly.

\section{Conclusion}
\bigskip

In summary, anomaly-induced 
masses are always present when supersymmetry is broken.
When these AMSB contributions dominate and
yield all the gaugino masses as well as adding to universal
scalar masses, a unique spectrum results which has important
differences from other models of supersymmetry.  
Several of these unorthodox features that 
arise in low-energy supersymmetry from 
the AMSB scenario include,
\begin{itemize}
\item The ratio of gaugino 
       masses $M_1:M_2:|M_3|$ is approximately $2.8:1:7.1$ when
      loop corrections are included, and $M_2\simeq m_Z$;
\item The lightest supersymmetric particle
      is most often the Wino, but may also be the sneutrino;
\item Dilepton signals with displaced vertices are 
       are useful signals for this scenario at LEP and
Tevatron;
\item The anomaly-induced contribution to  left and right slepton masses
      is accidentally degenerate.  This remains true if the 
      required, additional sources for slepton masses are universal;
\item LEP signatures are sensitive to the hierarchy of sneutrino, slepton 
      and Wino masses. 
      The searches in the different channels can be simply combined to give
      exclusion plots in the $M_2$-$m_{\tilde\nu_L}$
      plane;
\item  The gravitino mass is  much heavier than the masses of the
      other sparticles.  Consequently, the cosmological problem associated
      with gravitino decays during nucleosynthesis 
      is alleviated over much of parameter space;
\item In spite of its negligible thermal relic abundance, neutral Winos can
      form the galactic dark matter, since they are copiously produced,
      below their freeze-out temperature, from the primordial gravitino
      decays.
\end{itemize}
Discovery of several of the above
phenomenological implications is necessary to gain confidence that the AMSB
scenario is a proper description of nature.

\section*{Appendix}

Using Eqs.~(\ref{spectroscopy})-(\ref{spectroscopy3}), 
the anomaly-mediated spectrum is
\bea
\wompit{M_1} & = &  \frac{33}{5} \frac{g_1^2}{16\pi^2}  m_{3/2} \\
\wompit{M_2} & = & \frac{g_2^2}{16\pi^2} {m_{3/2}}\\
\wompit{M_3} & = & -3\frac{g^2_3}{16\pi^2}{m_{3/2}}\\
\womp{m^2_{\tilde t_R}} & = & 
 \left( -\frac{88}{25}g^4_1+8g_3^4 +2y_t\hat\beta_{y_t} 
  \right) \frac{m_{3/2}^2}{(16\pi^2)^2} \\
\womp{m^2_{\tilde b_R}} & = & 
  \left( -\frac{22}{25}g^4_1+8g_3^4+2y_b\hat\beta_{y_b} 
     \right) \frac{m_{3/2}^2}{(16\pi^2)^2}\\
\womp{m^2_{\tilde Q_3}} & = & \left( -\frac{11}{50}g^4_1-\frac{3}{2}g_2^4+8g^4_3
 +y_t\hat\beta_{y_t}+y_b\hat\beta_{y_b} \right) \frac{m_{3/2}^2}{(16\pi^2)^2} \\
\womp{m^2_{H_u}} & = & \left( -\frac{99}{50} g^4_1 -\frac{3}{2}g^4_2 
 +3y_t\hat\beta_{y_t} \right) \frac{m_{3/2}^2}{(16\pi^2)^2} \\
\womp{m^2_{H_d}} & = & \left( -\frac{99}{50}g^4_1 -\frac{3}{2}g^4_2 
 +3y_b\hat\beta_{y_b} +
       y_{\tau}\hat\beta_{y_\tau} \right) \frac{m_{3/2}^2}{(16\pi^2)^2} \\
\womp{m^2_{\tilde L_3}} & = & \left( -\frac{99}{50}g^4_1-\frac{3}{2}g_2^4
 +y_{\tau}\hat\beta_{y_\tau}\right) \frac{m_{3/2}^2}{(16\pi^2)^2} \\
\womp{m^2_{\tilde \tau_R}} & = & \left( -\frac{198}{25}g^4_1 +2y_\tau 
  \hat\beta_{y_\tau}\right) \frac{m_{3/2}^2}{(16\pi^2)^2} \\
 A_{y_t} & = & -\frac{\hat\beta_{y_t}}{y_t}\frac{m_{3/2}}{16\pi^2} \\
 A_{y_b} & = & -\frac{\hat\beta_{y_b}}{y_b}\frac{m_{3/2}}{16\pi^2} \\
 A_{y_\tau} & = & -\frac{\hat\beta_{y_\tau}}{y_\tau}
        \frac{m_{3/2}}{16\pi^2} .
\eea
where
\bea
\hat\beta_{y_t} & = & 16\pi^2\beta_{y_t} =y_t\left( -\frac{13}{15}g^2_1 -3g_2^2 
  -\frac{16}{3}g^2_3 +6y_t^2 +y_b^2 \right) \\
\hat\beta_{y_b} & = & 16\pi^2\beta_{y_b} =y_b\left( -\frac{7}{15}g^2_1-3g^2_2
  -\frac{16}{3}g^2_3 +y_t^2 +6y_b^2+y_\tau^2\right) \\
\hat\beta_{y_\tau} & = & 16\pi^2\beta_{y_\tau} =y_\tau \left( 
 -\frac{9}{5}g^2_1-3g^2_2+3y_b^2
 +4y_\tau^2 \right).
\eea
The first two generation squark and slepton masses are obtained by 
appropriately changing the Yukawa couplings to first and second generation
Yukawa couplings.


\end{document}